\documentclass[useAMS,usenatbib,usedcolumn]{mn2e}

\def\HI{\ifmmode{\rm HI}\else{H\/{\sc i}}\fi}

\def\lsun{\ifmmode{{\mathrm L}_{\odot}}\else{L$_{\odot}$}\fi}

\def\msun{\ifmmode{{\mathrm M}_{\odot}}\else{M$_{\odot}$}\fi} 
\def\msunpc2{\ifmmode{{\mathrm M}_{\odot} \, {\mathrm{pc}}^{-2}}\else{M$_{\odot} \, {\mathrm {pc}}^{-2}$}\fi}

\def\kms{\ifmmode{{\mathrm{km \, s^{-1}}}}\else{${\mathrm{km \, s^{-1}}}$}\fi}

\usepackage[figuresright]{rotating}
\usepackage{lscape}
\usepackage{graphics}
\usepackage{epsfig}
\usepackage{multirow}
\usepackage{bigdelim}
\usepackage{bigstrut}
\usepackage{amsmath}

\title[Rotation curves of S\'ersic bulges]{The rotation curves of flattened
  S\'ersic bulges}   
\author[Edo~Noordermeer]
  {Edo~Noordermeer\thanks{email:edo.noordermeer@nottingham.ac.uk} \\
  University of Nottingham, School of Physics and Astronomy, University
  Park, NG7 2RD Nottingham, UK}

\begin{document}

\date{accepted, 9-12-2007}

\maketitle

\begin{abstract}
  I present a method to deproject the observed intensity profile of an
  axisymmetric bulge with arbitrary flattening to derive the 3D luminosity
  density profile and to calculate the contribution of the bulge to the
  rotation curve. I show the rotation curves for a family of fiducial
  bulges with S\'ersic surface brightness profiles and with various
  concentrations and intrinsic axis ratios. Both parameters have a
  profound impact on the shape of the rotation curve. In particular, I
  show how the peak rotation velocity, as well as the radius where it
  is reached, depend on both parameters. 

  I also discuss the implications of the flattening
  of a bulge for the decomposition of a rotation curve and use the case of
  NGC~5533 to show the errors that result from neglecting it. For NGC~5533,
  neglecting the flattening of the bulge leads to an overestimate of its
  mass-to-light ratio by approximately 30\% and an underestimate of the
  contributions from the stellar disc and dark matter halo in the regions
  outside the bulge-dominated area.  
\end{abstract}


\section{Introduction}
\label{sec:introduction}
Rotation curves of spiral galaxies are one of the prime methods to probe the
mass distribution on galactic scales. 
The discovery, made in the 1970s \citep{Rogstad72,Roberts75,Bosma78,Bosma81b}, 
that rotation curves remain flat till the last measured points, well outside 
the optical discs, provided the decisive evidence for the presence of large 
amounts of dark matter in galaxies. 
Unfortunately, however, it has proven difficult to gain detailed knowledge
about the {\em distribution} of dark matter in galaxies. 
This is primarily due to the fact that, with the exception of the
lowest surface brightness galaxies, the stars in galaxies contribute a
significant, but not precisely known, fraction to the overall
gravitational potential
\citep{Persic96,DeBlok97,Palunas00,mijnproefschrift}. 

To determine the contribution from the stars to an observed rotation curve,
the full 3D structure of the stellar mass distribution needs to be known. 
In most disc galaxies, the bulk of the stars can be divided between a flat
disc and a thicker central bulge. 
Once an accurate bulge-disc decomposition has been performed, it is relatively 
straightforward to calculate the contribution to the rotation curve from the
disc, given its simple, effectively 2D structure \citep[e.g.][]{Casertano83}. 
For the central bulge, however, the situation is more complex. 
Historically, it was often assumed that bulges follow an $R^{1/4}$
luminosity-profile \citep{DeVaucouleurs58,VanHouten61,DeVaucouleurs74}.
For dynamical purposes, they were often further assumed to be spherically
symmetric bodies, in which case the observed intensity profile can be
deprojected and the gravitational potential calculated easily
\citep[e.g.][]{Young76}. 
In reality, however, bulges are generally neither spherical, nor do they
follow the $R^{1/4}$ density-profile
\citep{Andredakis94,Andredakis95,Erwin99,Graham01,Erwin02,Noordermeer07a}. 

Several papers have addressed various aspects related to the properties of
non-spherical bulges with a more general luminosity density profile. 
The problem of deprojecting a non-spherical bulge was already addressed by
\citet{Stark77}.
The dynamical properties of spherical bodies following a general $R^{1/n}$
luminosity profile have been studied by \citet{Ciotti91}. 
This analysis was extended by \citet{Trujillo02}, who studied triaxial systems
following the $R^{1/n}$ luminosity law and presented an analytical expression
to approximate the deprojected, 3D luminosity density profile. 

For the purpose of rotation curve analysis, one usually assumes axisymmetry
for all components in the galaxy and ignores the triaxial nature of
the bulge.
Most bulges are only mildly triaxial \citep{Bertola91,
Mendez-Abreu07}, so that the errors stemming from this assumption are
generally small.  
Accounting for the full triaxiality of the bulges merely complicates the 
calculations in this context.
In fact, given the fact that rotation curves are by definition
one-dimensional quantities, they are intrinsically unsuitable to
account for deviations from axisymmetry. 
A proper treatment of the dynamical effects of triaxial bulges
requires two- or three-dimensional modelling of the gas flows in their
potentials. 
The flattening of the bulge in the vertical direction, on the other hand, 
{\em is} often significant \citep[e.g.][]{Bertola91, Noordermeer07a}.
It is also dynamically important and should not be neglected, even in
the axisymmetric approximation.  

In this paper, I study the 3D luminosity density of axisymmetric 
bulges with various intrinsic flattenings and their contribution to the 
rotation curve of a galaxy. 
In section~\ref{sec:derivation}, I present the relevant equations to
deproject the observed surface brightness profile of a bulge to the 3D
luminosity density and to calculate its contribution to the rotation curve.  
Although some of the equations in section~\ref{sec:derivation} can be
applied to an arbitrary bulge surface brightness profile, I specifically focus
on bulges with a S\'ersic luminosity profile \citep{Sersic68}.  
In section~\ref{sec:properties}, I present the bulge rotation curves for a
range of parameters and discuss various properties of the curves. 
I show the effects of the concentration and flattening of the bulge on the
shape of its rotation curve and in particular on the location and height of
the peak.  
Finally, in section~\ref{sec:discussion}, I briefly discuss the implications
of the results from the preceding sections. 
In particular, I use the example of the Sab galaxy NGC~5533 to illustrate the
errors which result from neglecting the flattening of the bulge when
decomposing a rotation curve.

\section{Derivation of the rotation curves}
\label{sec:derivation}

\subsection{Deprojection of the observed surface density distribution}
\label{subsec:deprojection}
The first step in our calculations involves the deprojection of the observed
surface density distribution, to yield the 3D luminosity density. 
This has been done for the general triaxial situation by \citet{Stark77}. 
Here, I consider the simpler configuration of an axisymmetric spheroid,
symmetric with respect to the plane of the galactic disc. 

Let us define an $(x,y,z)$-coordinate system such that the $(x,y)$-plane
coincides with the plane of symmetry of the galaxy. 
The luminosity density can then be written as $\rho (x,y,z) = \rho(m)$,
with $m \equiv \sqrt{x^2 + y^2 + (z/q)^2}$ and $q$ the intrinsic axis ratio
of the bulge isodensity surfaces.
\begin{figure}
 \centerline{\psfig{figure=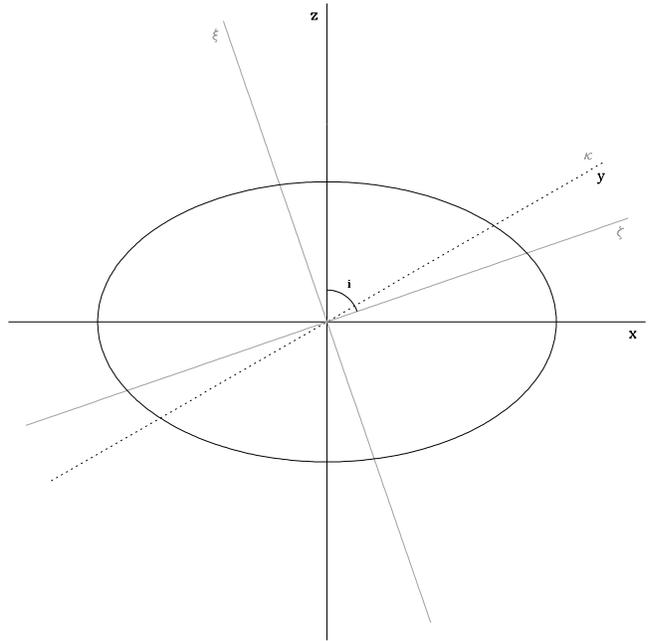,height=8.5cm}}
  \caption{Schematic overview of the two coordinate systems. Black
  lines give the galactic coordinate system, grey lines give the
  coordinate system of the observations (see text). The ellipse shows
  an $(x,z)$-cut through an isodensity surface. 
  \label{fig:coords}} 
\end{figure}  

A second coordinate system, $(\zeta,\kappa,\xi)$, can be defined corresponding
to the observational configuration. 
$\zeta$ lies along the line of sight and is inclined by an angle $i$ with
respect to the $z$-axis.  
The $\kappa$-axis coincides with the $y$-axis; both are defined to lie
along the line of nodes.  
$\xi$ is the axis that is perpendicular to $\kappa$ on the plane of the sky;
in practice this means that $\xi$ lies along the minor axis of the galaxy. 
See figure~\ref{fig:coords} for an overview of the coordinate systems.   
The $(x,y,z)$ and $(\zeta,\kappa,\xi)$ coordinate systems are related by the
following transformation matrices:
\begin{equation}
 \label{eq:coordtransa}
 \begin{array}{c@{\hspace{0.15cm}}c@{\hspace{0.15cm}}c@{\hspace{0.cm}}l} 
   \left( \begin{array}{c} \zeta \\ \kappa \\ \xi \end{array} \right) & 
    =                                                                  & 
   \left( \begin{array}{ccc} 
           \hspace{0.2cm}  \sin i & 0 & \hspace{0.2cm}\cos i \\
           \hspace{0.2cm}    0    & 1 & \hspace{0.2cm}  0    \\
           \hspace{-0.1cm}-\cos i & 0 & \hspace{0.2cm}\sin i \end{array} 
                                                               \right) &
   \left( \begin{array}{c} x \\ y \\ z \end{array}            \right)\mbox{ }
 \end{array}
\end{equation}
\begin{equation}
 \label{eq:coordtransb}
 \begin{array}{c@{\hspace{0.15cm}}c@{\hspace{0.15cm}}c@{\hspace{0.cm}}l} 
   \left( \begin{array}{c} x \\ y \\ z \end{array}            \right) & 
    =                                                                  & 
   \left( \begin{array}{ccc} 
           \hspace{0.18cm}\sin i & 0 & \hspace{-0.1cm}-\cos i \\
           \hspace{0.18cm}  0    & 1 & \hspace{0.2cm}   0     \\
           \hspace{0.18cm}\cos i & 0 & \hspace{0.2cm} \sin i \end{array} 
                                                               \right) &
   \left( \begin{array}{c} \zeta \\ \kappa \\ \xi \end{array} \right). 
 \end{array}
\end{equation}

With these coordinates, the projected intensity at each position 
$(\kappa,\xi)$ on the sky can be written, in the case of an optically thin
medium, as 
\begin{equation}
\label{eq:projection}
I(\kappa,\xi) = \; \int_{\substack{{} \\[0.3cm] \hspace{-0.35cm}
  \zeta=-\infty}}^{\substack{\hspace{-0.17cm} \vspace{0.25cm} \infty} \\ {}}
  \! \! \rho(\zeta,\kappa,\xi) \, d\zeta.
\end{equation}
It can be shown that the projected intensity distribution on the sky
has an ellipsoidal shape, where the ellipticity $\epsilon$ of the 
projected image is related to the intrinsic axis ratio $q$ and the
inclination angle $i$ as 
\begin{equation}
\label{eq:bulgeflattening}
(1-\epsilon)^2 = q^2 + (1-q^2) \cos^2 i 
\end{equation}
Thus, the problem is reduced to finding the relation between the
projected intensity at the line of nodes ($\xi=0$) and the density
$\rho(m)$. 
Once the intensity at the line of nodes is derived, the rest of the
image can be reconstructed using
equation~\ref{eq:bulgeflattening}. 

Given a position $\kappa_0$ at the line of nodes ($\xi=0$), the relation
between $\zeta$ and $m$ is (see
equations~\ref{eq:coordtransa}, \ref{eq:coordtransb}):  
\begin{eqnarray}
\label{eq:mvszeta_a} 
 m^2(\zeta \mid \kappa_0, \xi=0) \! \! & = & \! 
                                          x^2 (\zeta \mid \kappa_0, \xi=0) + 
                                          y^2 (\zeta \mid \kappa_0, \xi=0) +
                                                               \nonumber \\ 
                                 \! \! &   & \! 
                            \frac{1}{q^2} z^2 (\zeta \mid \kappa_0, \xi=0) 
                                                                         \\  
                                 \! \! & = & \!
                                          \zeta^2 \sin^2 i + \kappa_0^2 +
                                  \frac{\zeta^2 \cos^2 i}{q^2} \nonumber \\
\label{eq:mvszeta_b}
 \zeta^2(m \mid \kappa_0, \xi=0) \! \! & = & \! 
               \frac{m^2 - \kappa_0^2}{\sin^2 i + \frac{1}{q^2}\cos^2 i} \\ 
\label{eq:mvszeta_c}
 \frac{\partial \zeta}
      {\partial m}               \! \! & = & \! 
              \frac{m}{\sqrt{
                              \left( \sin^2 i + \frac{1}{q^2}\cos^2 i \right)  
                              \left( m^2 - \kappa_0^2 \right)
                            }}                           
\end{eqnarray}
Then, setting $\xi=0$ in equation~\ref{eq:projection}, changing the
integration variable from $\zeta$ to $m$ and using
equation~\ref{eq:mvszeta_c}, the projected intensity $I(\kappa_0)$ at 
the line of nodes becomes: 
\begin{eqnarray}
\label{eq:MAproj}
I(\kappa_0) & = & \hspace{0.235cm} 
                    \int_{-\infty}^{\infty} 
                       \rho(\zeta \mid \kappa_0, \xi=0) \, d\zeta 
                                                          \nonumber \\ 
            & = & 2 \int_{0}^{\infty} 
                       \rho\big(m(\zeta \mid \kappa_0, \xi=0)\big) \, d\zeta 
                                                          \nonumber \\[-0.2cm]
            &   &                                                   \\[-0.2cm]
            & = & 2 \int_{m(0 \mid \kappa_0, \xi=0)}
                        ^{m(\infty \mid \kappa_0, \xi=0)}
                       \rho(m) \frac{\partial \zeta}{\partial m} \, dm 
                                                          \nonumber\\  
            & = & \frac{2}{\sqrt{\sin^2 i + \frac{1}{q^2} \cos^2 i}}
                    \int_{\kappa_0}^{\infty} \rho(m) 
                        \frac{m \, dm}{\sqrt{m^2 - \kappa_0^2}}. 
                                                          \nonumber 
\end{eqnarray}
This is an Abel-integral, which can be inverted
\citep[e.g.][app.~1.B4]{Binney87} to obtain the following equation, which
gives the 3D luminosity density distribution $\rho(m)$ when the observed
major-axis intensity profile $I(\kappa)$ is given: 
\begin{equation}
\label{eq:rho3D}
\rho(m) = - \frac{1}{\pi} 
          \sqrt{\sin^2 i + \frac{1}{q^2}\cos^2 i} 
          \int_{m}^{\infty} \frac{dI}{d\kappa} 
                            \frac{d\kappa}{\sqrt{\kappa^2 - m^2}}. 
\end{equation}
\begin{figure*}
 \centerline{\psfig{figure=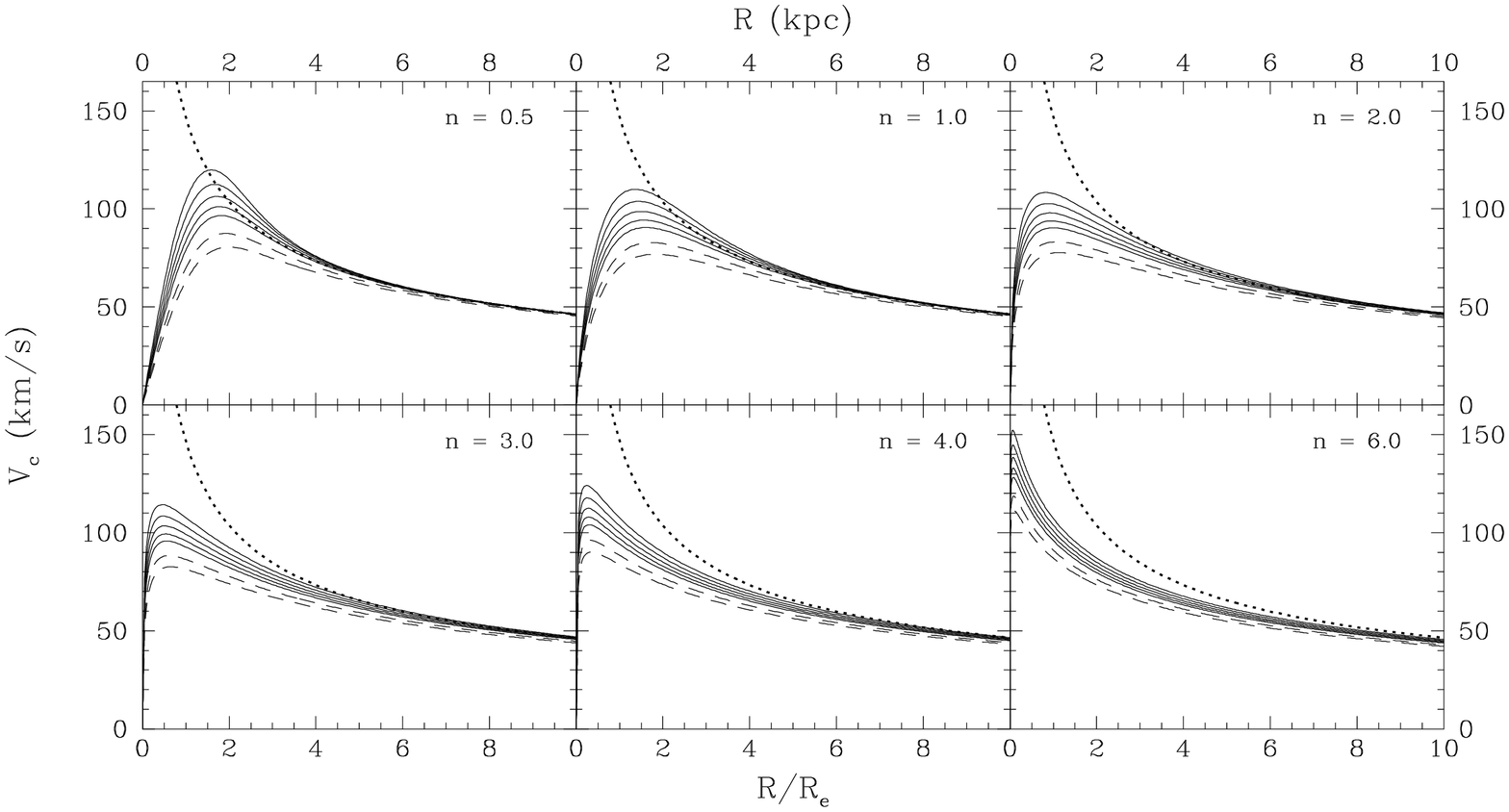,angle=270,width=17.cm}}
  \caption{Rotation curves for bulges with total mass $M_b = 5 \cdot 10^9 \,
    \msun$ and effective radius $r_e = 1 \, \mathrm{kpc}$. The different
    panels are for different values of the S\'ersic concentration parameter
    $n$, as indicated in the top-right corners. The solid lines show, from top
    to bottom, rotation curves for axis ratios $q = 0.2, 0.4, \ldots, 1.0$
    respectively. The dashed lines are for the unrealistic cases of prolate
    bulges ($q=1.5$ and $2.0$). The dotted lines show the Keplerian rotation
    curve for a point-like body of equal mass.  
  \label{fig:rotcurs}} 
 \centerline{\psfig{figure=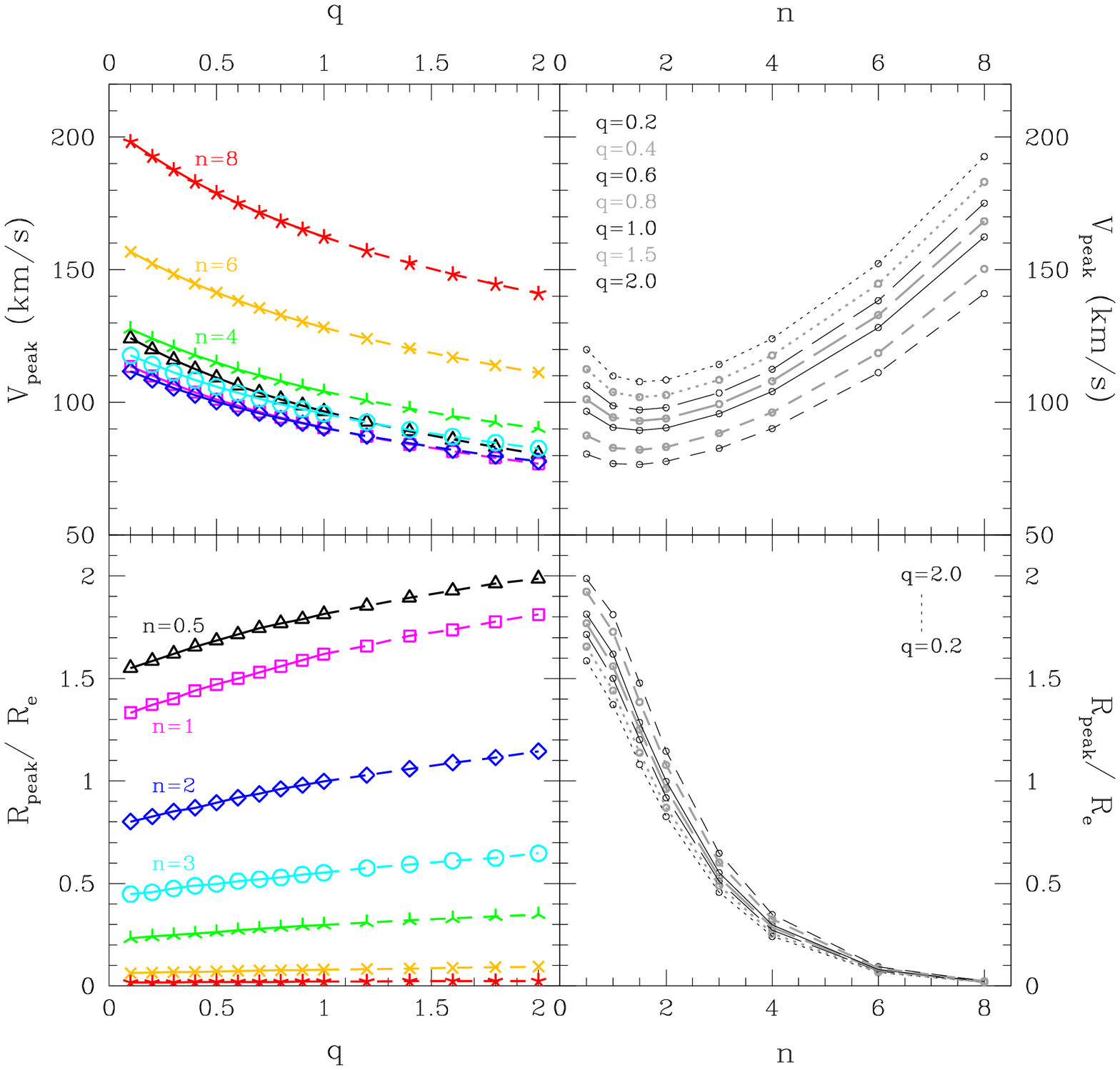,width=12.cm}}
  \caption{The dependence of the peak rotation velocity ($V_{\mathrm{peak}}$,
    top panels) and the corresponding radius ($R_{\mathrm{peak}}$, bottom
    panels) on the bulge flattening ($q$, left hand panels) and the
    concentration parameter ($n$, right hand panels). Different colours and
    symbols in the left hand panels indicate different values for $n$, as
    labeled; dashed lines indicate unrealistic values of $q$ corresponding to
    prolate bulges. The different lines in the right hand panels show the
    behaviour for different values of $q$. 
  \label{fig:maxrot}} 
\end{figure*}

\subsection{Calculation of the rotation curve}
\label{subsec:rotcurcalc}
Now, assuming that the observed emission traces mass and that the
mass-to-light ratio is constant throughout the bulge, equation~\ref{eq:rho3D}
also gives the 3D mass density distribution in the bulge (except for a
constant factor for the actual value of the mass-to-light ratio, $\Upsilon$).
The derivation of the corresponding rotation curve is then straightforward and
is described in \citet{Binney87}. 
Inserting equation~\ref{eq:rho3D} into their equation 2-91, we get for the
rotation curve: 
\begin{eqnarray}
\label{eq:bulgerotgen}
V^2(r) & = & - 4 G q \Upsilon \sqrt{\sin^2 i + \frac{1}{q^2}\cos^2 i} \;\;
                                                   \times \nonumber \\[-0.2cm]
       &   &                                                        \\[-0.2cm] 
       &   &     \int_{\substack{{} \\[0.25cm] \hspace{-0.3cm} m=0}} 
                     ^{\substack{\hspace{-0.12cm} \vspace{0.25cm} r}} 
                     \; \left[ \int_{\kappa=m}^{\infty} 
                                   \frac{dI}{d\kappa} 
                                   \frac{d\kappa}{\sqrt{\kappa^2 - m^2}}
                       \right] \frac{m^2 \, dm}{\sqrt{r^2 - m^2 e^2}}, 
                                                          \nonumber
\end{eqnarray}
where $e$ is the eccentricity of the bulge: $e = \sqrt{1 - q^2}$. 

This equation is valid for any given observed intensity profile $I(\kappa)$;
no prior assumptions have been made regarding the shape of the light or
density profiles, except that of spheroidal symmetry. 
In principle, one could directly derive the rotation curve for any given
intensity profile by numerically calculating the derivative $dI/d\kappa$
and evaluating the integral. 
In practice, the intensity profiles of many bulges follow a general S\'ersic
profile \citep{Sersic68}, defined as:
\begin{equation}
\label{eq:Sersic}
I_b(\kappa) = 
          I_{0} \exp \left[ - \left( \frac{\kappa}{r_0} \right)^{1/n} \right].
\end{equation}
Here, $I_0$ is the central surface brightness and $n$ is a concentration
parameter that describes the curvature of the profile in a radius-magnitude
plot. 
For $n=4$, equation~\ref{eq:Sersic} reduces to the well-known de Vaucouleurs
profile \citep{DeVaucouleurs48}, whereas $n=1$ describes a simple exponential
profile. 
$r_0$ is the characteristic radius, which is related to the effective radius
($r_e$, the radius which encompasses 50\% of the light) as $r_e = b_n^{\; n}
\, r_0$. 
$b_n$ is a scaling constant that is defined such that it satisfies
$\gamma(b_n, 2n) = \frac{1}{2} \Gamma(2n)$, with $\gamma$ and $\Gamma$ the 
incomplete and complete gamma functions respectively. 
Given a S\'ersic bulge, one can evaluate the derivative $dI/d\kappa$
analytically:  
\begin{equation}
\label{eq:dSersic}
\frac{dI}{d\kappa} = - \frac{I_0}{r_0 n} \exp 
                       \left[ - \left( \frac{\kappa}{r_0} \right)^{1/n} \right]
                    \: \left( \frac{\kappa}{r_0} \right)^{1/n - 1}.
\end{equation}

Inserting this into equation~\ref{eq:bulgerotgen}, we get the following, final
equation for the bulge rotation curve: 
\begin{eqnarray}
\label{eq:bulgerot_app} 
V_b^2(r) & = & {\cal C} 
               \int_{\substack{{} \\[0.25cm] \hspace{-0.3cm} m=0}}
                   ^{\substack{\hspace{-0.12cm} \vspace{0.25cm} r}} \;
                   \left[ 
               \int_{\kappa=m}^{\infty} 
                   \frac{e^{-(\kappa/r_0)^{1/n}} (\kappa/r_0)^{1/n - 1}}  
                        {\sqrt{\kappa^2 - m^2}} \: d\kappa 
                  \right] \; \; \times                     \nonumber \\[-0.2cm]
         &   &                                                       \\[-0.2cm]
         &   &     \frac{m^2}{\sqrt{r^2 - m^2 e^2}} \: dm, \nonumber \\[0.5cm]
{\cal C} & = & \frac{4 G q \Upsilon I_0}{r_0 n} \sqrt{\sin^2 i +
                                         \frac{1}{q^2}\cos^2 i}. 
\end{eqnarray}

\section{Properties of the rotation curves}
\label{sec:properties}
Equation~\ref{eq:bulgerot_app} can, to my knowledge, generally not be solved
analytically.  
Instead, it was evaluated numerically, using algorithms from
\citet{numrep}\footnote{A {\sc {FORTRAN}} program to evaluate the integrals in
  equation~\ref{eq:bulgerot_app} is available upon request from the author.}, 
for various parameter combinations.

Of particular interest are the effects of the flattening, $q$, and the
concentration, $n$, of the bulges on the resulting rotation curves. 
In figure~\ref{fig:rotcurs}, I show the rotation curves for a family of
fiducial bulges with various concentrations and axis ratios, but with $I_0$
and $r_0$ scaled such that all bulges have equal total luminosity ($L_b = 5
\cdot 10^9 \, \lsun$) and effective radius ($r_e = 1 \, \mathrm{kpc}$). For
simplicity, I assumed a mass-to-light ratio of $\Upsilon = 1$. 
The range of values for $n$ and $q$ which I consider here spans the observed  
values in bulges of nearby spiral galaxies of various morphological types 
\citep{Andredakis95,Khosroshahi00,Graham01,Noordermeer07a}, although I also
include a number of prolate bulges with $q > 1$ which are not observed in
reality.  
Note that observed bulges show a general trend of increasing luminosity and
radius with increasing $n$. 
For easy comparison, however, I choose to study bulges with equal luminosity
and effective radii here; the rotation curves shown in
figure~\ref{fig:rotcurs} can be scaled easily for bulges with different mass
or radius. 

It is clear that both the flattening and the concentration of the bulge have
a large influence on the shape of the rotation curves.
Varying either parameter changes both the radius where the rotation curve
peaks and the corresponding peak velocity. 
This is illustrated in a quantitative way in figure~\ref{fig:maxrot}. 
As expected, a flattened bulge has a higher peak rotational velocity than a
spherical one. 
Interestingly, the strength of the correlation is a weak function of the
concentration, being largest for less concentrated bulges.
For $n=0.5$, a highly flattened bulge with an intrinsic axis ratio of $q=0.25$
has a peak velocity about 22\% higher than the spherical case. 
For a highly concentrated bulge ($n = 8$), the difference is approximately
17\%. 

Not surprising either, more centrally concentrated bulges reach the peak in
their rotation curve at smaller radii. 
Cored bulges with $n=0.5$ reach the maximum rotation velocity between 1.5 and
2.0 effective radii, depending on $q$. 
Concentrated bulges with $n=8$, on the other hand, peak at very small radii,
within 2\% of the effective radius, with the exact radius again depending on
$q$. 

The dependence of the location of the peak in the rotation curve on the
flattening of the bulge is somewhat less expected. 
The bottom left hand panel in figure~\ref{fig:maxrot} shows that the peak is
reached at smaller radii in more flattened bulges. 
The strength of the correlation depends on the concentration of the bulge: for
$n=0.5$, a highly flattened bulge with an intrinsic axis ratio of $q=0.25$
reaches the peak rotation velocity at a 12\% smaller radius than the spherical
case, whereas for $n=8$, this difference is 18\%.
As an aside, I note that a spherical bulge ($q=1$) with a S\'ersic parameter
of $n=2$ reaches the peak rotational velocity at exactly one effective
radius.

Finally, for a given total mass, effective radius and intrinsic axis ratio,
the peak rotational velocity is lowest for a bulge with $n=1.5$, with both
more and less concentrated bulges having higher peak rotation velocities. 
\begin{figure*}
 \centerline{\psfig{figure=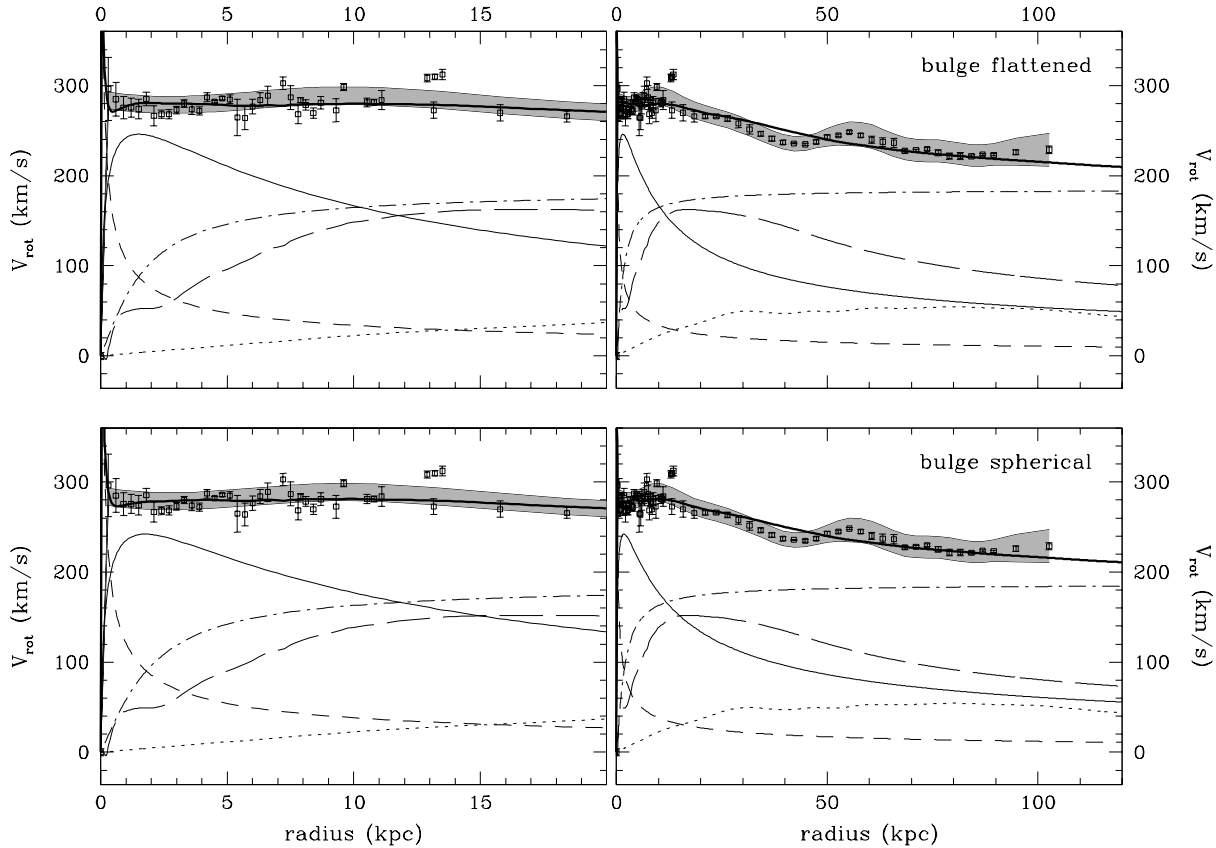,width=16cm}}
  \caption{Rotation curve decomposition for NGC~5533. Top panels show the
    decomposition using the correct value of the bulge flattening. The bottom
    panels show the case where the bulge flattening was neglected and its
    contribution was calculated under the assumption of spherical
    symmetry. Left hand panels show the central regions while the right hand
    panels show the full rotation curve. The data points, errorbars and gray
    shaded band show the observed rotation curve and uncertainties
    (see \citealt{Noordermeer07b} for details). The thin lines show the
    contributions from bulge (solid), stellar disc (long dashed), gaseous disc
    (dotted), central black hole (short dashed) and dark matter halo
    (dot-dashed). The bold line shows the rotation curve for the combined
    model.
  \label{fig:ugc9133}} 
\end{figure*}

\section{Discussion}
\label{sec:discussion}
It is now common knowledge that morphologically, the assumptions that bulges
are spherical and have an $R^{1/4}$ luminosity-profile are poor descriptions
of reality. 
Our analysis shows that this is also true dynamically.
As figures~\ref{fig:rotcurs} and \ref{fig:maxrot} illustrate, the flattening
and concentration of a bulge have a strong influence on the shape of its
rotation curve. 
Specifically, non-spherical bulges and bulges with a concentration parameter
$n$ different from 4 have rotation curves with very different peak velocities
and radii than a `classical' spherical de Vaucouleurs bulge. 

The dependence of the peak velocity on the flattening of the bulge has
important consequences in the decomposition of an observed rotation curve into
contributions from the various components in a galaxy.
In practice, the bulge contribution in such a decomposition is often
determined by scaling up the bulge mass-to-light ratio until the peak of the
bulge rotation curve matches the amplitude of the observed curve at that
radius.
If we neglect the intrinsic flattening of a bulge and instead derive its
rotation curve under the assumption that it is spherically symmetric, we will
underestimate the peak velocity (see figure~\ref{fig:rotcurs}). 
As a result, we will need to assume a higher mass-to-light ratio in order to
explain the observed rotation velocity at the peak radius. 
Figure~\ref{fig:maxrot} shows that the difference in peak velocity between a
spherical bulge and a highly flattened bulge with $q = 0.25$ is of the order
of 20\%. 
Since the mass-to-light ratio is proportional to the square of the bulge
rotation velocity (equation~\ref{eq:bulgerot_app}), the error in the former
would be more than 40\%.   

To illustrate this effect, I show in figure~\ref{fig:ugc9133} the rotation
curve decomposition for the massive Sab galaxy NGC~5533.  
Its rotation curve was measured from a combination of optical spectroscopic
data and \HI\ observations by \citet{Noordermeer07b}. 
R-band photometry for this galaxy was presented by \citet{Noordermeer07a}, as
well as a detailed study of the properties of its bulge and disc. 
The effects of the seeing on the measured bulge parameters were
corrected for, using the deconvolutions from \citet{Graham01}. 
The bulge was found to be highly flattened, $q = 0.3$, making this galaxy an
interesting test-case for the effect described above. 

Based on the derived properties of the bulge from \citet{Noordermeer07a}, I
have derived two versions of its contribution to the rotation curve: one
accounting correctly for its flattening using equation~\ref{eq:bulgerot_app}
and another by neglecting the flattening and assuming $q=1$. 
The contribution of the stellar and gaseous discs were calculated using the
prescriptions from \citet{Casertano83}, assuming a vertical scale height of
one fifth of the stellar disc scale length. 
For simplicity, the dark matter halo was assumed to have an isothermal density
profile.
Finally, in order to explain the rapid rise of the rotation curve in the
centre, a central black hole was added to the model. 
The contributions from the various components were then added and the 5 free
model parameters (the R-band mass-to-light ratios $\Upsilon_b$
and $\Upsilon_d$ for the bulge and disc respectively, the core radius $R_c$
and central density $\rho_0$ of the dark halo and the mass of the black hole
$M_{BH}$) were adapted in a simple $\chi^2$ minimisation procedure to obtain
the best fitting model rotation curve.  
The values for the fitted parameters are listed in table~\ref{table:ugc9133}.  
\begin{table}
 \centering
  \caption[Rotation curve decomposition of NGC~5533.]
   {Results from the rotation curve decomposition of NGC~5533. The middle
   column shows the fitted parameters for the case where the flattening of the
   bulge was taken into account, while the right-hand column is for the
   situation where the bulge was assumed to be spherically
   symmetric. The errors are the formal fitting uncertainties, and do
   not include any systematic uncertainties, nor the degeneracies
   between the various parameters. \label{table:ugc9133}}  
   \begin{tabular}{lr@{$\pm$}lr@{$\pm$}l}
    \hline
     & \multicolumn{2}{c}{$q=0.33$} & \multicolumn{2}{c}{$q=1$} \\
    \hline
    $\Upsilon_b$ ($\msun/L_{\odot,R}$)           & 2.8  & 0.1  & 3.7  & 0.1  \\
    $\Upsilon_d$ ($\msun/L_{\odot,R}$)           & 5.0  & 0.2  & 4.4  & 0.2  \\
    $R_c$        (${\mathrm{kpc}}$)              & 1.4  & 0.2  & 1.7  & 0.2  \\
    $\rho_0$     ($\msun \, {\mathrm{pc}}^{-3}$) & 0.31 & 0.08 & 0.23 & 0.05 \\
    $M_{BH}$     ($10^9 \msun$)                  & 2.7  & 0.6  & 3.4  & 0.6  \\
    \hline    
   \end{tabular}
\end{table}

Figure~\ref{fig:ugc9133} shows that equally good fits can be made with both
bulge rotation curves; neglecting the bulge flattening does not have a
noticeable effect on the quality of the model in this case. 
However, it is clear from table~\ref{table:ugc9133} that, when the bulge is
assumed to be spherically symmetric, a higher mass-to-light ratio has to be
assumed to explain the observed rotation velocities around the peak radius of
2~kpc.  
In this case, the difference between the two values for $\Upsilon_b$ is 32\%. 

Figure~\ref{fig:ugc9133} and table~\ref{table:ugc9133} also illustrate a more
subtle effect.  
The effect of the flattening of the bulge on its rotation curve contribution
diminishes rapidly with radius (see also figure~\ref{fig:rotcurs}). 
Outside about 5 effective radii, the flattening has virtually no effect
anymore on the amplitude of the bulge rotation curve, which then only depends
on the total mass of the bulge.
Thus, because we were forced to assume a higher mass-to-light ratio for the
spherical bulge, its contribution at large radii is larger than in the
flattened case.
As a result, the derived contribution of the disc is somewhat smaller and the
dark matter halo is less concentrated, with a 21\% larger core radius. 
The same effect occurs when we assume an NFW density profile
\citep{Navarro97} for the dark matter halo.  
In that case, we find again that the concentration of the halo is
significantly smaller (by about 15\%) when we assume that the bulge is
spherical, compared to the model where we account for the flattening
of the bulge.

In conclusion, the results shown in this paper prove that, for a detailed
study of the dynamical structure of disc galaxies with a sizeable bulge, it is
crucial to properly account for the concentration and intrinsic flattening of
the latter.

\section*{Acknowledgements}
I would like to thank Mike Merrifield for stimulating discussions and helpful
suggestions during the preparations of this paper. I am also grateful
to the referee, Enrico Corsini, for useful suggestions. 

\bibliographystyle{mn2e}
\bibliography{../../references/abbrev,../../references/refs}

\end{document}